\documentclass{iaus}
\usepackage{graphicx}

\title[Hadronic jet models] 
{Hadronic jet models today}

\author[Marek Sikora]   
{Marek Sikora$^1$}

\affiliation{$^1$Nicolaus Copernicus Astronomical Center, \\ 
Bartycka 18, 00-716 Warsaw, Poland \\ 
email: {\tt sikora@camk.edu.pl} 
}

\pubyear{2010}
\volume{275}  
\pagerange{1--8}
\jname{Jets at all Scales}
\editors{G.E. Romero, R.A. Sunyaev \& T. Belloni, eds.}
\begin{document}

\maketitle

\begin{abstract}
The matter content of relativistic jets in AGNs is dominated by a 
mixture of protons, electrons, and positrons.  During dissipative events
these particles tap a significant portion of the internal and/or kinetic 
energy of the jet and convert it into electromagnetic radiation.
While leptons -- even those with only mildly relativistic energies - can  
radiate efficiently, protons need to be accelerated up to 
energies exceeding $10^{16-19}$ eV to dissipate radiatively a significant amount
of energy via either trigerring pair cascades  or direct 
synchrotron emission.   
Here I review various constraints imposed on the role of 
hadronic non-adaiabatic cooling processes in
shaping the high energy  spectra of blazars. It will be argued that protons, 
despite being efficiently accelerated and presumably playing a crucial role
in jet dynamics and dissipation of the jet kinetic energy to the internal
energy of electrons and positrons, are more likely to remain radiatively 
passive in AGN jets.

\keywords{galaxies: active, (galaxies:) BL Lacertae objects: general,
 galaxies: jets, (galaxies:) quasars: general, gamma rays: theory
}
\end{abstract}

\firstsection 
\section{Introduction}
Production of relativistic jets in AGN is very likely mediated by
rotation of large scale magnetic fields in magnetosphere of black hole and/or 
accretion disk (\cite[Blandford 1976]{Blandford76};
\cite[Lovelace 1976]{Lovelace76}; \cite[Phinney 1983]{Phinney83};
\cite[Camenzind 1986]{Camenzind86}; \cite[McKinney \& Blandford 2009]{kb09}). 
This leads to a formation of Poynting flux dominated outflows. Those in turn 
can at some point be  converted to the matter dominated jets 
(\cite[Komissarov et al. 2007]{KBVK07}; \cite[Tchekhovskoy et al. 2009]{tkn09};
\cite[Lyubarsky 2010]{Lyubarsky10}; \cite[Komissarov 2010]{kom10}), 
with a terminal Lorentz factor $\Gamma \sim P_j /\dot M c^2$ 
where $P_j$ is the rate of energy extraction from rotating BH and/or 
accretion disk and $\dot M$ is the mass flux. Depending on whether 
a jet is launched in the BH magnetosphere or by the accretion disk, the mass 
flux is expected to be dominated by electron/positron pairs or by protons.

Presence of protons in AGN jets is  indicated by 
the low-energy cutoffs in the radio spectra of hot spots in radio-lobes
(\cite[Blundell et al. 2006]{Blun06}; \cite[Stawarz et al. 2007]{Staw07};
\cite[Godfrey et al. 2009]{God09}) 
and by circular polarization and Faraday Rotation  in radio cores 
(\cite[Vitrishchak et al. 2008]{Vitri08}; 
\cite[Park \& Blackman 2010]{PB10}). 
If so,  such protons might manifest their presence also via
radiative contribution to high energy spectra in blazars, by synchrotron
emission of pair cascades triggered by photo-meson process and by direct 
synchrotron emission of protons and mesons
(\cite[Mannheim \& Biermann 1992]{mb92}; \cite[Mannheim 1993]{Mannheim93};
\cite[Rachen \& M\'esz\'aros 1998]{rm98}; \cite[Aharonian 2000]{Aharonian00}).
Whether such a contribution can be significant is the main issue of 
this presentation. We start with 
a short review of the leading hadronic models (\S2); discuss efficiencies of
the proton acceleration and cooling processes (\S3);  
present observational constraints on hadronic contribution to blazar spectra
(\S4). And, finally, we discuss the role of protons they play in the leptonic
models regarding the aspect of preheating electrons up 
to thermal proton energies  required to participate 
in further stochastic 
acceleration process (\S5). Main results are summarized in \S6.

\section{Hadronic models - basic features}

\smallskip
\centerline{2.1. {\it Luminous blazars - 'photo-meson' models}}
\smallskip

Luminous blazars are hosted by quasars which, when observed away from
the axis of the jet, 
form population of FRII type double radio sources. They are powered by 
accretion onto BHs with masses typically  of the order of $10^9$ solar masses
and their accretion luminosities are in the range 
$10^{46} - 10^{47} \rm{erg~s^{-1}}$. However, when such jets are 
oriented close to the line of sight, these quasars are seen as being
 dominated by  nonthermal radiation of jets,  with the apparent
luminosities $10^{48-49} {\rm erg~s^{-1}}$ and spectra
dominated by the broad high energy component peaking in the 1-100 MeV 
band. This radiation shows high amplitude variability on time scales from 
years down to days and even hours, implying the strong dissipative events 
taking place not far from the base of a jet. There the energy densities of 
the magnetic and radiation fields  are very large, providing conditions for 
acceleration of protons up to energies of 
$~ 10^{9-10}$ GeV, and for cooling them via inelastic collisions
with soft photons (\cite[Sikora et al. 1987]{Sik87}). For typical background radiation fields  
most of proton energy is converted to mesons and this initiates processes 
which according to proposers
of hadronic models are responsible for $\gamma$-ray production in luminous
blazars (\cite[Mannheim \& Biermann 1992]{MB92}). 

These processes are dominated by the following channels.
In approximately 90\% collisions, the produced mesons are pions.
In 2/3 of them they are the neural pions ($\pi^0$) and in 1/3 of them  
--  the positive pions $\pi^{+}$.
Neutral pions  almost immediately decay into photons
which in turn trigger pair-cascades driven by photon-photon 
pair production and their synchrotron radiation. 
Escaping  radiation is the product of 3rd and 4th generation
of pairs (radiation of the  first two are totally 
converted to $e^+e^-$-pairs). The resulting  
electromagnetic spectrum is predicted to form 
the high energy component peaked in the $\gamma$-ray band, with a high energy
break at $h\nu_{br} \sim 10-30$ GeV, where $\tau_{\gamma\gamma \to e^+e^-} \simeq 1$,
and low energy tail in the X-ray band
with a slope $\alpha_X > 0.5$.
 
Such spectra can be affected, but not significantly,
by electromagnetic output resulting from production of charged pions via
processes $p\gamma \to n \, + \pi^{+}$ and $n\gamma \to p \, + \pi^-$.
The charged pions decay producing muons 
($\pi^{\pm} \to \mu^{\pm} + \bar \nu_{\mu}/\nu_{\mu}$), and the muons decay
producing positrons/electrons ($\mu^{\pm} \to e^{\pm} + \bar\nu_e/\nu_e + 
\nu_{\mu}/\bar\nu_{\mu}$). The resulting electrons/positrons join 
the pair cascade triggered by the decays of the neutral pions and together
with the synchrotron emission of muons 
(\cite[Rachen \& M\'esz\'aros 1998]{RM98}) contribute an additional
$\sim 6$-$16$ \% to the electromagnetic output of the photo-meson
process.
\smallskip

\centerline {2.2. {\it Low luminosity TeV BL Lac objects - proton-synchrotron 
models}}
\smallskip

The low luminosity BL Lac objects are hosted by radio-galaxies of type FR I.
Just as FR II radio sources, they are associated with giant elliptical galaxies,
with central BH masses  of the order of $10^9 {\rm M}_{\odot}$, but with
the jet powers at least 3 orders of magnitude lower than in the radio-loud quasars and extremely low
accretion luminosities.
Doppler boosted nonthermal radiation from their jets reaches TeV energies,
and the luminosity peak of the high energy spectral component is located
at GeV energies. Resulting from the low radiative environment, energy losses 
of ultra-relativistic protons in theses objects  are presumably dominated 
by direct synchrotron emission of protons. This mechanism was suggested 
by \cite[Aharonian (2000)]{Aha00} to be the primary source of 
$\gamma$-rays in low luminosity BL Lac objects. Main spectral  features of 
such models, as is in the case
of synchrotron radiation of electrons,  are directly related
to the magnetic field intensities and the injection function of relativistic
particles (here of protons).  
They produce photons with average energies 
$ \nu_{p,syn} = (2e/(3\pi m_pc)) \, \gamma_p^2 \, B' {\cal D}$
and energy spectra with the slopes $\alpha = (q_p-1)/2$ in the slow 
cooling regime and $\alpha = q_p/2$ in the fast cooling regime,
where $B'$ is the intensity of magnetic field in the jet co-moving frame,
${\cal D}$ is the Doppler factor, $q_p$ is the index of the proton
power-law injection function, $Q_p \propto \gamma_p^{-q_p}$, and $\alpha$ is 
the index of the radiation energy flux, $F_\nu \propto \nu^{-\alpha}$.
Application of those formulae indicates that  protons accelerated up to Lorentz 
factors  $\gamma_p \sim 10^{10}$
and cooled in magnetic fields $B' \sim 100$ Gauss may produce 
spectra reaching  TeV energies and with the luminosity peak 
located near $\nu_{p,syn, max}$.  
An investigation whether these parameters are feasible -- given the constraints imposed on 
$\gamma_{p,max}$ by jets with the limited power and magnetisation -- is given in \S{4.2}. 

\section{Radiative efficiencies}

\centerline{3.1.{\it Time scales}}
\smallskip

\noindent
{\underline{\it Acceleration}}.
 
Time scale of the proton acceleration as measured in the co-moving 
frame of the flow is
\begin{equation} t_{acc}' = f R_L/c = 
{m_p c \over e} \, {f \gamma_p \over B'} \, , \end{equation}
where $R_L$ is the Larmor radius, $B'$ is the magnetic field intensity, 
and $f$ is the parameter which in the case of shock acceleration depends on the  spectrum
of magnetic turbulence and on the velocity of the upstream-flow 
(\cite[Rieger et al. 2007]{Rieger07}) and 
for mildly relativistic shocks is expected
to be at least of the order of 10.
\smallskip

\noindent
{\underline{\it Adiabatic losses}} 

Assuming that cross-sectional radius of the source is of the order of the 
cross-sectional radius of the jet, $R$, relativistic plasma moving down the jet
with a Lorentz factor $\Gamma$ undergoes energy losses due to adiabatic
expansion. For conical jets with the half-opening angle 
$\theta_j \equiv R/r < 1/\Gamma$, where $r$ is the distance of the source 
in a jet, the time scale of the adiabatic
losses is (\cite[Moderski et al. 2003]{Mod03})
\begin{equation} t_{ad}' ={3 \over 2}\,  {R \over  (\theta_j\Gamma)c} \, .
\end{equation}
\noindent
{\underline{\it Photo-meson process}}

Time scale of the energy losses via the photo-meson process can be
estimated using the approximate formula (\cite[Begelman et al. 1990]{Beg90}) 
\begin{equation} t_{p\gamma}' \sim {1 \over  
\langle\sigma_{p\gamma}K_{p\gamma}\rangle c n_{ph}'(\nu'>\nu_{th}')} \,
, \end{equation}
where  
$\langle\sigma_{p\gamma}K_{p\gamma}\rangle \sim 0.7 \times 10^{-28} {\rm cm^2}$
is the product of the photo-meson cross-section and inelasticity parameter 
averaged over the resonant energy range,
$ n_{ph}'(\nu'>\nu_{th}') = \int_{\nu_{th}'}{ n_{\nu}'\, {\rm d} n_{\nu}'}$,
$h\nu_{th}' \simeq m_{\pi}c^2 / \gamma_p$ is the threshold photon
energy and $m_{\pi}$ is the rest mass of the pion.
The target radiation field is provided by the internal and external sources.
The internal seed soft radiation is dominated by synchrotron emission of 
primary (directly accelerated) electrons, the external one -- by 
re-scattered/reprocessed disk radiation.
Approximating the broad synchrotron spectral component by a power-law 
function with the energy-flux index $\alpha = 1$ and denoting its luminosity
by $L_s$, we have
\begin{equation} n_{ph(int)}'(\nu'>\nu_{th}') \sim  
{L_s \over 4 \pi m_{\pi} c^3 R^2 {\cal D}^4} \, \gamma_p
 \, . \end{equation}

Spectra of external radiation fields  are in turn narrow and therefore can be
approximated by mono-energetic functions. Hence,  
\begin{equation} n_{ph(ext)}' \sim {\xi L_d \Gamma \over 4 \pi c r^2  \, h \nu_{ext}} =
 {\xi L_d (\theta_j\Gamma)^2 \over 4 \pi c R^2 \Gamma \, h\nu_{ext}} \,\,
{\rm for}\, \gamma_p > m_{\pi}c^2/h\nu_{ext}  \, ,  \end{equation}
and $n_{ph,ext}'=0$ for $\gamma_p < m_{\pi}c^2/(\Gamma h\nu_{ext})$,
where $L_d$ is the luminosity of the accretion disk and $\xi$ is the fraction
of this luminosity rescattered/reprocesssed on a spatial scale corresponding
with a distance $r$ of the source in a jet. 
\smallskip

\noindent
{\underline{\it Synchrotron emission}}

Time scale of the proton cooling via the synchrotron process is 
\begin{equation} t_{p,syn}' = {3\over 4} \, \left( m_p \over m_e \right)^3 \, 
{m_e c \over \sigma_T u_B'} \, {1 \over \gamma_p} \, ,\end{equation}
where $u_B'= {B'}^2/(8\pi)$ is the magnetic energy density. 
\smallskip

\centerline{3.2.{\it 'Mono-energetic' efficiencies}}
\smallskip
In order to illustrate efficiencies  of the proton acceleration 
and cooling processes we introduce their dimensionless rates, as scaled by 
the adiabatic losses rates, $\tau_i \equiv {t'}_i^{-1}/{t'}_{ad}$.
They are:
\begin{equation}
 \tau_{acc} \simeq 
4.8 \times 10^{-7} \, {B' R \over f (\theta_j \Gamma) \gamma_p} \simeq \,
0.8 \, {(\sigma/0.1)^{1/2} L_{j,47}^{1/2} \over 
(f/10)\, (\Gamma/10) \, (\theta_j\Gamma)} \, {1\over \gamma_{p,10}} \, , \end{equation}
\begin{equation} \tau_{p\gamma}^{(int)} \sim 1.2 \times 10^{-36} \,
{L_s \,\gamma_p \over (\theta_j\Gamma)\, {\cal D}^4 \,R}
\, \simeq  12 {L_{s,47} \over 
(\theta_j \Gamma) \, (\Gamma/10)^4 R_{16}}\, \gamma_{p,10}
\, , \end{equation}
\begin{equation} \tau_{p\gamma}^{(ext)} \sim 2.8 \times 10^{-40} \,
{ \xi L_d \, (\theta_j \Gamma) \over h\nu_{ext} \, \Gamma R} \,
\simeq 0.2 \, { (\xi L_d)_{45}\, (\theta_j \Gamma) 
\over (h\nu_{ext}/10{\rm eV}) \, (\Gamma/10) \, R_{16}} \, , \end{equation}
and
\begin{equation} \tau_{p,syn} =  1.1 \times 10^{-29} \,
{R {B'}^2 \, \gamma_p \over \theta_j\Gamma} \simeq \,
0.3 \, {(\sigma/0.1) L_{j,47} \over 
(\theta_j\Gamma) \, (\Gamma/10)^2 \, R_{16}} \, \gamma_{p,10} \, , \end{equation}
where $L_B \simeq c u_B' \pi R^2 \Gamma^2 = \sigma L_j/(1+\sigma)$
is the magnetic energy flux,
$L_j= L_M + L_B$ is the total jet power, $\sigma \equiv L_B/L_M$, and $L_M$
is the energy flux of the rest mass. Assuming $\sigma <1$ we use 
approximation $L_B \sim \sigma L_j$. 

The 'scaled' quantities, $L_{j,47}$, $\gamma_{p,10}$, $R_{16}$, 
$L_{s,47}$, and $(\xi L_d)_{46}$ are defined in the usual way, 
i.e. $X_n \equiv X/10^n$. 
\smallskip

\centerline{3.3.{\it Maximal proton energies}}
\smallskip

Maximal proton energies -- if limited only by adiabatic losses -- can be found
from $\tau_{acc}=1$ to be
\begin{equation} \gamma_p(\tau_{acc}=1) \simeq 8 \times 10^{9} \,
{(\sigma/0.1)^{1/2} L_{j,47}^{1/2} \over (f/10)\, (\Gamma/10)\, (\theta_j\Gamma)} 
\, , \end{equation}
Stronger limits are imposed  if dominant energy losses are non-adiabatic,
i.e. for $\tau_{cool} = \tau_{p\gamma} + \tau_{p,syn} > 1$.
For energy losses dominated by photo-meson process with the target radiation
field provided by internal sources or by proton-synchrotron radiation
\begin{equation}  \gamma_{p,max} = \gamma_p(\tau_{acc}=1) \, {\rm Min}[1; \sqrt{R/R_c}] \, 
.\end{equation}
where in the 1st case (photo-meson process)
\begin{equation} R_c^{(p\gamma)} \simeq 9.5 \times 10^{16} \,
{L_{s,47} \, (\sigma/0.1)^{1/2} L_{j,47}^{1/2} \over
(\theta_j \Gamma)\, (\Gamma/10)^5 \, (f/10)} \,\, {\rm [cm]} . \end{equation}
and in the 2nd case (proton-synchrotron radiation)
\begin{equation} R_c^{(syn)} \simeq 2.3 \times 10^{15} \,
{(\sigma/0.1)^{3/2} L_{j,47}^{3/2}\,  \over 
(f/10) \, (\theta_j\Gamma)\, (\Gamma/10)^3 } \,\, {\rm [cm]}\, .\end{equation}  
For energy losses dominated by photo-meson process with the target radiation
field provided by external sources, $\tau_{p\gamma}^{(ext)}$ is predicted
to be lower than unity for any proton energy and therefore $\gamma_{p,max}$
is not expected to be affected. 
\smallskip

\centerline{3.4.{\it Total efficiencies}}
\smallskip

Radiative efficiency of a given cooling process can be estimated using
formula 
\begin{equation} \eta_i \simeq {\int_1^{\gamma_{p,max}} {
\eta_i(\gamma_p) Q_{\gamma_p}\gamma_p \, d\gamma_p} \over
\int_1^{\gamma_{p,max}} {Q_{\gamma_p}\gamma_p \, d\gamma_p} } \, , \end{equation}
where $\eta_i(\gamma_p) \simeq {\rm Min}[\tau_i; 1]$ and
$Q_{\gamma_p}$ is the proton injection function.
\smallskip

We calculate such efficiencies below  assuming power-law injection of protons
$Q_{\gamma_p} \propto \gamma_p^{-q_p}$ with  $q_p=2$. This specific value
of the index is chosen because efficiencies obtained
for $q_p=2$ provide  upper limits of efficiencies
available for $q_p>2$ being predicted by theoretical models of a diffusive 
acceleration of particles in mildly relativistic shocks and supported by 
the slopes of synchrotron spectra produced in the IR-Optical bands by primary 
(directly) accelerated electrons/positrons. 

For energy losses dominated by photo-meson process with internally produced
seed photons and by proton-synchrotron emission, we obtain
\begin{equation} \eta_i \simeq {1 + \ln{(\gamma_{p,max} /\gamma_{p1})}
\over \ln{\gamma_{p,max}} } \, {\rm for~} R<R_c, \end{equation}
and $\eta_i < 1/\ln{\gamma_{p,max}}$ for $R>R_c$, where
in case of 'internal' photo-meson process, $R_c$ is given by Eq.(13) and 
\begin{equation} \gamma_{p1}^{(p\gamma)} = 
\gamma_p(\tau_{p\gamma}^{(int)}=1) \simeq 8.3 \times 10^8 \,
{(\theta_j \Gamma))\, (\Gamma/10)^4 \, R_{16} \over L_{s,47}}
\, , \end{equation} 
while in case of proton-synchrotron emission, $R_c$ is given by Eq.(14) and
\begin{equation} \gamma_{p1}^{(syn)} = \gamma_p(\tau_{p,syn}) \simeq 
3.4 \times 10^{10} \, {(\theta_j\Gamma)\, (\Gamma/10)^2 \, R_{16} \over
(\sigma/0.1) L_{j,47} } \, . \end{equation}

For  energy losses dominated by photo-meson process with externally produced
seed photons, at any distance larger than 
\begin{equation} r = {\Gamma R(\tau_{p,\gamma}^{(ext)}=1) \over (\theta_j\Gamma)}\,  
\simeq 1.8 \times 10^{16} \,
{(h\nu_{ext}/10{\rm eV}) \, (\Gamma/10)^2 \over (\xi L_d)_{45} \, 
(\theta_j \Gamma)^2} \,\, {\rm [cm]}  \end{equation}
$\tau_{p,\gamma}^{(ext)} \le 1$ and the efficiency is
\begin{equation}\eta_{p\gamma}^{(ext)} \simeq \tau_{p\gamma}^{(ext)} \,
{\ln{(\gamma_{p,max}/\gamma_{p,th})} \over \ln\gamma_{p,max}} \, ,\end{equation}
where $\gamma_{p,th} = m_{\pi} c^2/ (\Gamma h\nu_{ext}) \simeq 
1.4 \times 10^6 / ((\Gamma/10)(h\nu_{ext}/10{\rm eV}))$.

\medskip

\section{Observational constraints}
\smallskip

\centerline{4.1.{\it Luminous blazars}}
\smallskip

In order to account for the $\gamma$-ray luminosities of powerful blazars,
radiative efficiency has to be
\begin{equation} \eta_i > 0.3 \, {L_{\gamma,48} \over (\Gamma/10)^2 \, (\eta_{diss}/0.3) \, 
L_{j,47}} \, , \end{equation}
where $\eta_{diss}$ is the fraction of the jet energy flux dissipated 
in the 'blazar zone'. As it can be verified using approximate
formulae presented  in \S{3.4}, such efficiency
is difficult to reach  even  for protons injected with the energy 
distribution slopes $q_p = 2$. For our fiducial parameters it
is about 10 \% for the photo-meson process with intenally produced
seed photons and much less for others.
For slope $q_p \sim 2.4$ the efficiency drops to $\eta \sim 10^{-3}$. 
\smallskip

Hadronic models may have also problems to explain very hard X-ray spectra of 
luminous blazars. Those blazars often have slopes $\alpha_x < 0.5$ 
(see Table 1 in \cite[Sikora et al. 2009]{Sik09}) and in order to explain 
such spectra by synchrotron
radiation of secondary $e^\pm$ -- products of the cascades
powered by hadrons -- one needs to assume 
inefficient cooling of ultra-relativistic electrons/positrons up to 
energies 
\begin{equation} \gamma_e > 8.0 \times 10^5 {(h\nu/100{\rm keV})^{1/2} \over
B' \, (\Gamma/10)} \, . \end{equation}
Inefficient cooling of such energetic electrons/positrons implies very weak
 magnetic 
fields and, therefore, puts strong constraints on the efficiency of the proton
acceleration and on efficiency of photo-meson energy losses via limitation
of the maximal proton energy. Furthermore, as Sikora et al. (2009)
demonstrated,  in order to avoid overproduction of X-rays in 
these magnetically weak sources by SSC radiation of primary electrons, it is 
necessary to assume the source sizes of the order of parsecs, and for such 
sources the efficiency of the photo-meson process is further reduced.
\smallskip

\centerline{4.2.{\it  Low luminosity BL Lac objects}}
\smallskip

In these  objects, because of low radiation energy densities -- both in jets 
themselves and in the surroundings of the jet -- the non-adiabatic energy 
losses of protons are presumably dominated by the proton-synchrotron mechanism.
However, noting that such objects 
are hosted by weak radio-galaxies, with the jet
powers $L_j \le 10^{44} {\rm erg~s^{-1}}$,  efficiency of
synchrotron-proton models  is also expected to be strongly reduced 
because weaker magnetic fields.  
In order to keep them at reasonable level more compact sources
must be assumed. However, even in the case of  most 
relativistic protons, the required  size of the source to provide sufficiently 
strong magnetic
fields for  efficient cooling is unreasonably small (see Eq. (3.14),
\begin{equation} R_c \sim 0.7 \times 10^{11} \, {(\sigma/0.1)^{3/2} L_{j,44}^{3/2} 
\over (f/10)\, (\theta_j \Gamma) \,(\Gamma/10)^3 } \, {\rm [cm]} . \end{equation}
This is 3 orders less than the gravitational radius of the BH with mass
$M_{BH} \sim 10^9 M_{\odot}$. Considering the minimal cross-sectional radius
of a jet to be $R \sim 10^{15}$cm, which for $\theta_j \sim1/\Gamma$ 
corresponds with a distance $~100$ gravitational radii of 
the $10^9 M_{\odot}$ BH -- required to be at least of this order to accelerate 
the jet up to $\Gamma \sim 10$ (\cite[Komissarov et al. 2007]{KBVK07}) --
we can find using Eqs. (3.10) and (3.11) that 
\begin{equation} \tau_{p,syn}(\gamma_{p,max}) \simeq 0.7 \times 10^{-4} \,
{(\sigma/0.1) L_{j,44})^{3/2} \over (f/1) \, (\theta_j\Gamma) \, (\Gamma/10)^3 \,
R_{15}} \, . \end{equation}
This indicates that even for such extreme parameters as $f \sim 1$ and 
$\sigma \sim 1$ the efficiency is too small to explain 
$\gamma$-ray luminosities $L_{\gamma} \sim 10^{44}{\rm erg~s^{-1}}$,
unless one assumes very hard ($q_p < 1$) proton injection spectra
and adopts significantly larger total jet power. 
\smallskip

The main purpose of proton-synchrotron models was to explain the relatively 
stable shape of the TeV spectra in variable low luminosity 
BL Lac objects (\cite[Aharonian 2000]{Aha00}). Obviously, the critical issue of such models is 
whether protons can reach sufficiently large energies to produce synchrotron
spectra extending up to TeV energies. Combining formulae for average 
synchrotron photon energies, magnetic energy flux, and  maximal proton 
energies (Eq. 3.11) gives 
\begin{equation} h\nu_{p,syn, max} \simeq  
3  \times 10^{-5} \, {(\sigma/0.1)^{3/2} L_{j,44}^{3/2} \over
(f/10)^2 \, (\Gamma/10)^2 \, (\theta_j\Gamma)^2 \,  R_{15}} \, 
[{\rm TeV}] . \end{equation}
One can see that spectra may extend to TeV energies only if assuming 
$f\sim 1$, $\sigma \sim 1$, and jet powers 
$L_j > 10^{45}{\rm erg~s^{-1}}$.

\section{Protons in leptonic models}

Disproving interpretation of high energy spectra of blazars produced via 
hadronic models does not disprove the presence of protons
in AGN jets.  They simply are expected to be radiatively inefficient but 
are likely to dominate the jet energy flux and strongly
affect the dynamics of dissipation processes via shocks in the regions
of low magnetization parameter ($\sigma < 0.1$). In this scenario,
a large fraction of dissipated energy must be converted to relativistic
electrons, otherwise leptonic  models will be inefficient despite 
high radiative efficiencies of energetic electrons. However in order
to get electrons to participate together with 
protons in the diffusive shock acceleration process,
electrons must be first preheated up to average energy (to be strict --
up to average momentum) of protons heated by
randomization and compression in the shocked plasma. Several scenarios
have been investigated for such electron preheating, both analytically and in 
PIC-simulations
(\cite[Amato \& Arons 2006]{AA06}; \cite[Amano \& Hoshino 2009]{AH09};
\cite[Sironi \& Spitkovsky 2010]{ss10}). 
Some of them indicate the formation of  a power-law energy distribution 
with the slope $1<q_e<2$. If it is true, then the number of electrons
joining protons in the stochastic acceleration process can be lower than 
the number of protons by significant factor. Hence, the high efficiency
of blazar radiation may indicate that $q_e <1$, or that there is a significant
pair content.

In the ERC (External-Radiation-Compton) models 
(\cite[Sikora et al. 1994]{Sik94}) of $\gamma$-ray production
in luminous blazars  such a low-energy tail
should be observed in the 30keV--1 MeV spectral energy range.
Unfortunately, these bands are observationally  poorly covered, particularly
above 20 keV. A number of blazars have been detected  up to $\sim 50 $keV,  
by INTEGRAL (\cite[Beckmann et al. 2009]{BRS09}), by 
Swift/BAT (\cite[Ghisellini et al. 2010]{Ghi10}), and by
others (see Table 1 in  \cite[Sikora et al. 2007]{sik07}). 
Some of them have spectral slopes 
corresponding with $1<q_e<2$. 
This, together with the location of the break in the energy range $1-10$MeV 
as suggested by CGRO/COMPTEL and OSSE observations
(\cite[Zhang et al. 2005]{Zhang05};
\cite[McNaron-Brown et al. 1995]{NB95}), may indicate a moderate 
($n_e/n_p \sim 10$) pair content, implied also by studies of the bulk-Compton 
and Compton-rocket effects (\cite[Sikora \& Madejski 2000]{SM00}; 
\cite[Ghisellini \& Tavecchio 2010]{GT10}). At energies $<$ 20 keV
  most blazars have softer X-ray spectra, but they can result
from the contribution of the  SSC process and/or
from the superposition of X-rays produced at different locations
in a jet, such as the orphan X-ray outburst detected in 3C279 
(\cite[Abdo et al. 2010]{Abdo10}).

\section{Conclusions}

For realistic jet powers ($L_j \le L_{Edd}$ in luminous blazars and
$L_j \le 10^{-3}L_{Edd}$ in low luminosity BL Lac objects), limited magnetization 
($\sigma \le 0.1$ is required to allow formation of strong shocks), 
and moderately steep proton injection function ($q_p \ge 2$ is suggested
by IR-optical spectra and by theoretical acceleration models), the 
hadronic models {\it fail to}:

\noindent
$\bullet$  
reproduce $\gamma$-ray luminosities of blazars;

\noindent
$\bullet$
explain formation of very hard X-ray spectra in luminous blazars;

\noindent
$\bullet$
provide the spectral extension up to TeV energies in low luminosity blazars.
\smallskip

Nevertheless, as indicated by several independent observations, protons
{\it are} present in AGN jets and presumably  play key role in 
dissipation processes in shocks.
In particular, they transfer a fraction of the dissipated energy 
to electrons/positrons helping them to reach threshold energies for
further acceleration by stochastic mechanisms and produce the 
observed $\gamma$-ray spectra via the ERC and SSC scenarios.  
However, significant pair content may be required to achieve reasonable
effiecincy of that energy transfer.

\bigskip

\noindent
{\bf Acknowledgements}

I thank G. Madejski, K. Nalewajko and {\L}. Stawarz for helpful comments.
The work was supported by Polish grant MNiSW NN203 301635.

\bigskip
\bigskip
\bigskip\bigskip\bigskip

\noindent
{\bf Discussion}
\smallskip

\noindent
BEDNAREK: Can curvature energy losses of protons be important in AGNs?

\smallskip
\noindent
SIKORA: They cannot be. This is because following particle acceleration
processes in shocks or reconnection regions, relativistic protons are 
injected with a broad distribution of pitch angles. Such protons (as well
as all other charged particles) spiral around magnetic field lines rather than
slide on them. 

\smallskip
\noindent
YUAN: There are two kinds of B field in a jet. One is a large scale helical
field, another being turbulent in a shock front. When we calculate radiation 
spectrum, we use turbulent one, but when we calculate polarization, we seem
to use the large scale ordered field.

\smallskip
\noindent
SIKORA: Properly calculated polarization must take into account both,
the tangled/turbulent magnetic fields compressed and/or generated  in a shock, 
as well, as the large scale magnetic fields transmitted through the shock.
 
\smallskip
\noindent
DERMER: Acceleration to high energies is faster for ions than protons. 
How do your conclusions change if you consider Fe rather than p?

\smallskip
\noindent
SIKORA: For approximately solar abundances of AGN plasmas the number of heavy 
nuclei is $\gg Z$ times smaller than of protons. Therefore, despite the fact 
that heavy nuclei are accelerated  $Z$ times faster, they contribute to 
radiative processes much less than protons and our main conclusion
that hadronic models cannot reproduce $\gamma$-ray luminosities of blazars
remains valid.

\smallskip
\noindent
PIRAN: Auger indicates that UHECRs are nuclei. It is much easier to calculate
nuclei. Nuclei will diffuse in the intergalactic field and can propagate
only up to $\sim 10$ Mpc. This suggestes that Cen A is the main source of
UHECRs if those are nuclei and if the intergalactic magnetic field is 
$> 10^{-9}$ G (Piran et al. 2010).  

\smallskip
\noindent
SIKORA: Yes, I agree that Cen A may contribute significantly to the observed
UHECRs.

\end{document}